\documentclass[aps,showpacs,preprint,preprintnumbers,amsmath,amssymb]{revtex4-1}

\usepackage{mathrsfs}
\usepackage{lineno}
\usepackage{amsmath}
\usepackage{amssymb}
\usepackage{graphicx}
\usepackage{color}
\usepackage{dcolumn}
\usepackage{bm}

\begin{document}

\title{Angle-dependent magnetoresistance and its implications for Lifshitz transition in W$_2$As$_3$}

\author{Jialu Wang$^{1,4}$, Haiyang Yang$^{1}$, Linchao Ding$^{2}$, Wei You$^{1}$, Chuanying Xi$^{3}$, Jie Cheng
$^{5}$, Zhixiang Shi$^{6}$, Chao Cao$^{1}$, Yongkang Luo$^{2,*}$, Zengwei Zhu$^{2}$, Jianhui Dai$^{1,*}$, Mingliang Tian$^{3,*}$, Yuke Li$^{1,*}$}

\affiliation{
$^{1}$ Department of Physics and Hangzhou Key Laboratory of Quantum Matters, Hangzhou Normal University, Hangzhou 311121, China\\
$^{2}$ Wuhan National High Magnetic Field Center and School of Physics, Huazhong University of Science and Technology, Wuhan 430074, China\\
$^{3}$ High Magnetic Field Laboratory, Chinese Academy of Sciences, Hefei 230031, China\\
$^{4}$ School of Science, Westlake Institute of for AdvancedStudy, Westlake University, Hangzhou 310024, China\\
$^{5}$ College of Science, Center of Advanced Functional Ceramics, Nanjing University of Posts and Telecommunications, Nanjing, Jiangsu 210023, China\\
$^{6}$ School of Physics and Key Laboratory of MEMS of the Ministry of Education, Southeast University, Nanjing 211189, China
}

\date{}

\begin{abstract}
Lifshitz transition represents a sudden reconstruction of Fermi surface structure, giving rise to anomalies in electronic properties of materials. Such a transition does not necessarily rely on symmetry-breaking and thus is topological. It holds a key to understand the origin of many exotic
quantum phenomena, for example the mechanism of extremely large magnetoresistance (MR) in topological Dirac/Weyl semimetals. Here, we report studies of the angle-dependent MR (ADMR) and the thermoelectric effect in W$_2$As$_3$ single crystal. The compound shows a large unsaturated MR (of about 70000\% at 4.2 K and 53 T). The most striking finding is that the ADMR significantly deforms from the horizontal dumbbell-like shape above 40 K to the vertical lotus-like pattern below 30 K. The window of 30-40 K also corresponds substantial changes in Hall effect, thermopower and Nernst coefficient, implying an abrupt change of Fermi surface topology. Such a temperature-induced Lifshitz transition results in a compensation of electron-hole transport and the large MR as well. We thus suggest that the similar method can be applicable in detecting a Fermi-surface change of a variety of quantum states when a direct Fermi-surface measurement is not possible.

\end{abstract}

\maketitle

\section{\label{sec:level1}Introduction}

A Lifshitz transition means an abrupt change of the Fermi surface topology of metals without symmetry breaking\cite{Lifshitz}, therefore, it is also termed as electronic topological transition. Conventional Lifshitz transition is \textit{quantum} phase transition at zero temperature, driven by chemical doping, magnetic field, pressure or uniaxial stress, in the vicinity of which the topology of the Fermi surface deforms \cite{Lifhitzdoping,li2017pressure,LifMag,Steppke:2017}. Indeed, such Lifshitz transition is associated with a variety of emergent quantum phenomena like van-Hove singulary, non-Fermi-liquid behavior, unconventional superconductivity \textit{et. al.}.  A recent famous example is Sr$_2$RuO$_4$, whose $\gamma$-Fermi sheet hits the edge of Brillouine zone and forms a van-Hove singularity under uniaxial stress. The critical transition temperature increases from $\sim$1.5 K up to $\sim$3.5 K \cite{Steppke:2017}, and meanwhile transport, magnetic and thermodynamic properties change correspondingly\cite{Barber-PRL2018,LuoY-Sr2RuO4NMR}. Besides those quantum control parameters, temperature can be regarded as another driving force to induce a Lifshitz transition, because the chemical potential ($\mu_F$) is temperature-dependence and its shift can modify the Fermi surface structure\cite{Ashcroft-SSP}. However, such realistic examples are rare \cite{WTe2Lifshize,ZhouXJZrTe5,zhou2016hall} due to the stringent requirements: (i) small Fermi energy ($\varepsilon_F$ in the order of $k_BT$$\lesssim$100 meV), so that the varying temperature can be influential to $\mu_F$ with respect to $\varepsilon_F$; (ii) band structure near the Fermi energy displaying anomalous dispersion, so that the contoured Fermi surface experiences an abrupt change.

Recently, a wide spectrum of topological semimetals have been extensively studied including Dirac semimetals (DSM)\cite{wang2012dirac,liu2014discovery,jeon2014landau}, Weyl semimetals (WSM)\cite{weng2015weyl,Lv-TaAsNP,Lv-TaAsPRX,xu2015discovery,shekhar2015extremely,ghimire2015}
as well as nodal-line semimetals\cite{FangC-NLSM,BianG-PbTaSe2,Madhab-ZrSiSTNLS}. These topological semimetals have bearing on a common feature: the extremely large magnetoresistance (MR)\cite{ali2014large,LuoY-WTe2Hall,tafti2015LaSb,li2016resistivity,YKLuo,Sjiang,zhou2016hall}, here the MR is defined as the percentage of the change of resistance in an applied magnetic field. Unlike previously colossal/giant MR observed in some magnetic materials, these newly discovered topological semimetals are generally nonmagnetic. Different mechanisms have been proposed to be responsible for the extremely large MR, including conventional electron-hole compensation\cite{ali2014large}, topological protection\cite{liang2015ultrahigh}, open-orbit effect\cite{Pippard1989}, and quantum limit\cite{abrikosov1998quantum}. Therefore what is the origin of MR in these semimetals is still under debate. In addition, these compounds usually hold of small $\varepsilon_F$ and complicated band structure near Fermi level, and hence set up a natural platform to search for new examples of temperature-induced Lifshitz transition.

On the other hand, the Fermi surface in materials plays a key role in unveiling the mechanism of MR, because its topology translates into
charge carriers trajectory, and hence electrical conductivities, in magnetic fields. For some topological semimetals with multi-band Fermi sheets, their Fermi surfaces are very complex and actually material-dependence, and the extrinsic control parameters can tune the deformation of Fermi surface topology. Therefore a direct Fermi-surface measurement is usually very hard and high-intensive. At this point, the measurements of angular dependent magnetoresistance (ADMR), as a simple and effective method to probe the information of Fermi surface or the possible topological phase transition in these materials, are highly required.

In this work, we first report a temperature-induced Lifshitz transition between 30 and 40 K in a candidate topological semimetal W$_2$As$_3$ by ADMR measurements. In the window of temperature the extremely large MR starts to appear correspondingly. We speculate that a downward shift of chemical potential as $T$ decreases has caused the emergence of a new hole pocket below 30 K, which fulfills the electron-hole compensation and leads to the extremely large MR. All of Hall effect, thermopower and Nernst effect measurements further lend support for this conclusion. We thus demonstrate the important role of electron-hole compensation to the extremely large MR in such kind of topological semimetal. Apart from this, we also suggest that the simple ADMR measurements can be applicable in detecting a Fermi-surface change in a variety of quantum states at a low expense, especially when a direct Fermi-surface measurement is not possible.

\section{\label{sec:level1}Results and discussions}

Before displaying our experimental results, it is necessary to briefly elucidate the ADMR as an effective technique to detect the Fermi surface anisotropy. For more details, the readers may refer to the Ref.\cite{NPong1994,ShengNZhang2019}. In uncorrelated compounds, MR typically originates from the bending of carrier trajectory. Watching in the momentum space, electrons (or holes) perform orbits about Fermi surface cross-sections perpendicular to $\mathbf{B}$. Considering a simple case, an ellipsoidal Fermi surface elongates along $\mathbf{k_z}$. For $\mathbf{B}\parallel\mathbf{k_z}$, the Fermi surface cross-section perpendicular to $\mathbf{B}$ is a circle. The scattering path length vector, defined as $\mathbf{l_k}=\mathbf{v_k}\tau$ (where $\mathbf{v_k}=\frac{1}{\hbar}\nabla_{\mathbf{k}}\varepsilon_{\mathbf{k}}$), sweeps out a circle in the $\mathbf{l}$ space. Whereas for $\mathbf{B}\parallel\mathbf{k_x}$, the Fermi surface cross-section now is an ellipse with variable local curvature $\kappa$. The swept out $\mathbf{l}$ curve also becomes non-uniform. The anisotropic shape of the Fermi surface causes different cyclotron masses and $\mathbf{v_k}$ (which depends on the local curvature of FS) under various field orientations, which has an effect on the ease of trajectory bending under magnetic fields and the value of MR as well. The situation becomes more interesting in a semimetal with multiple Fermi surfaces, as the anisotropy now further affects the degree of carrier compensation and leads to strongly angular dependent MR\cite{ShengNZhang2019}. ADMR, thus, is deemed as a powerful method to reflect the change of Fermi surface in transports. The Fermi surface topology can be more complicated in realistic materials, e.g., multiple Fermi surface pockets, open Fermi surface, or Fermi surface with anomalous curvatures. Such ADMR measurements, however, are still useful to unveil the Fermi surface characteristics. A typical example is the conventional semimetal Bismuth\cite{NPZhuBi,PhysRevXBi}.

The compound studied in this work, W$_2$As$_3$, was originally synthesized over a half century ago\cite{W2As3}. It crystalizes in a monoclinic structure C2/$m$ (No. 12) with the angle spanned by $\mathbf{a}$ and $\mathbf{c}$ being $\beta$=124.7$\textordmasculine$. For many decades, the physical properties of W$_2$As$_3$ have remained unclear until recently, Li et al reported the large MR and the Shubnikov de Hass (SdH) quantum oscillations\cite{PhysRevB.98.115145ZAXu}. The topological nature was also suggested by first-principles calculation. Besides that, the abnormal transport properties in this system, such as angular dependent magnetoresistance (ADMR) and thermoelectric effect, which are associated with the anisotropy of Fermi surface and the possible topological phase transition, are really of absence.

Figure 2(a) displays the longitudinal resistivity ($\rho\equiv\rho_{xx}$) of W$_2$As$_3$ as a
function of temperature ($T$) down to 2 K with the electrical current $\bf{I}\parallel\bf{b}$ and the applied
magnetic field $\bf{B}\parallel \bf{a}$ (cf. Fig.3(b)). For $B$=0, $\rho_{xx}(T)$, displays highly metallic behavior, attains a magnitude of 315 $\mu \Omega\cdot$cm at room temperature but falls to 1.04 $\mu \Omega\cdot$cm (=$\rho_0$) at 2 K,
yielding a large residual resistivity ratio RRR$=$303, comparable with previous finding\cite{PhysRevB.98.115145ZAXu}. The $\rho_{xx}(T)$ decreases sublinearly for $T$$>$80 K, and gradually evolves into a power-law for $T$$<$40 K . We write down the formula $\rho_{xx}$=$\rho_0$$+$$AT^n$, and display the results of $\log(\rho_{xx}-\rho_{0})$ vs. $\log(T)$ in the inset of Fig.2(a), giving rise to the exponent $n$=3.32(6).
The $n$ value accidently falls in between $n =$ 2 for $e-e$ scattering and $n =$ 5 for electron-phonon ($e-ph$)
scattering process according to the Bloch-Gruneisen theory\cite{ziman1960}, and is very close to $n =$ 3 and 4 for LaBi and LaSb\cite{sun2016large,tafti2015LaSb}.
The application of a magnetic field does not change the resistivity too much in the high temperature region, but enhances it substantially below $T_m$$\approx$30-40 K. The origin of $T_m$ will be discussed in detail later on. For low fields, $\rho_{xx}(T)$ tends to saturate at low temperatures; as $B$ increases, the field-induced upturn in $\rho_{xx}(T)$ becomes more and more prominent, resulting in the large magnetoresistance MR$\equiv$$(\rho(B)-\rho(0))/\rho(0)$. Similar behavior has been observed in many other topological semimetals\cite{ali2014large,xu2015discovery,tafti2015LaSb}.

The MR as a function of $B$ for several selected temperatures is shown in Fig. 2(b). The positive MR with a parabolic field-dependence reaches a value of 2300\% at 9 T and 2 K. The MR decreases rapidly with increasing temperature and finally holds less than a few percentages at 200 K, analogous to the most of topological semimetals. To verify the unsaturated MR at the higher magnetic field, we perform the pulsed-field measurements up to 53 T at 4.2 K, as shown in the inset to Fig. 2(b). The magnitude of MR reaches about 70000\% at 53 T without any signature of saturation. Different to the Dirac semi-metal Cd$_3$As$_2$ with the linear MR\cite{liang2015ultrahigh}, here the MR follows a nearly perfect parabolic field-dependence, implying the perfectly compensated electron and hole system predicted by the two-band theory\cite{ziman1960,ali2014large},
\begin{equation}
\rho_{xx}(B)=\frac{1}{e}\frac{(n_h\mu_h+n_e\mu_e)+(n_h\mu_e+n_e\mu_h)\mu_h\mu_eB^2}{(n_h\mu_h+n_e\mu_e)^2+(n_h-n_e)^2
\mu_h^2\mu_e^2B^2},
\label{Eq1}
\end{equation}
where $n_e$ ($n_h$) and $\mu_e$ ($\mu_h$) are electron (hole) carrier densities and mobilities, respectively. The equality of $n_e$ and $n_h$ gives rise to exact $B^2$ law of $\rho_{xx}(B)$, therefore, the compensated semimetals usually exhibit the large MR, such as in WTe$_2$, TaAs$_2$ and LaSbTe\cite{ali2014large,YKLuo,singha2016low}. In addition, the increase of carrier mobilities (see figure 4c) in these semimetals is also beneficial to the large MR, because the MR ($\propto \mu_h \mu_eB^2$) is proportional to the mobility at low temperatures. The wiggles on top of the parabolic background stemming from the SdH oscillations will be discussed further hereafter.

Another striking finding in our sample W$_2$As$_3$ is the deformation of ADMR below $T_m$. The geometry of our measurement is depicted in Fig. 3(b). The field-rotation is about $\mathbf{b}$-axis and thus is kept to be perpendicular to electric current. One can clearly see the ADMR, and more interestingly, these patterns change dramatically with temperature. In Fig. 3(c-i) we show the ADMR in polar plots at selected temperatures between 2 and 50 K. Above 35 K, the ADMR exhibits strong twofold symmetry at 9 T [Fig.3(c)], highly consistent with the configuration of
centro-symmetry crystal structure in W$_2$As$_3$. This dumbbell-like pattern suggests anisotropy Fermi surface.
The maximum of resistance occurs at $\theta=0$ ($\mathbf{B}\parallel\mathbf{a}$), and the minimum localizes at $\theta=90\textordmasculine$, while no clear feature can be identified at around $\theta=\beta$ (i.e. $\mathbf{B}\parallel\mathbf{c}$). As $T$ decreases to 34 K, four extra lobes show up around $n\pi/3$ ($n$=$\pm$1,$\pm$2). Further decreasing temperature, the resistivity for $\theta=90\textordmasculine$ grows very rapidly, and  finally this takes over the maximum [Fig. 3(e-h)]. The similar feature retains down to 2 K. Over the full temperature range 2-50 K, the resistivity for $\theta=0$ keeps a local maximum although a very shallow dip emerges in the vicinity of $\mathbf{B}\parallel\mathbf{a}$ at very low temperatures. Whereas, what attracts us most is that the pattern of ADMR changes from a horizontal dumbbell-like to a vertical lotus-like, in the temperature range from 40 K to 30 K. This reminds us of the Lifshitz transition of which the topology of Fermi surface undergoes a substantial reconstruction.

Hall effect and thermoelectric effect are very sensitive
to the electronic structure of sample. We present the results of Hall resistivity ($\rho_{yx}$), thermopower ($S$), and Nernst coefficient ($\upsilon$) in Fig. 4. The Hall resistivity,
as shown in Fig. 4(a), is essentially negative for all the measured temperatures (2-100 K) and fields (0-9 T), manifesting that electrons are the majority carrier. For $T>$70 K, $\rho_{yx}(B)$ is essentially linear with $B$. It starts to
decrease gradually and bends strongly at high magnetic fields below 50 K.
As T$ \leq$ 30 K, a positive slope in $\rho_{yx}$(B) above $\sim$ 6 T is observed. The slope
extends gradually to low magnetic fields and becomes more prominent until 15 K, where
$\rho_{yx}$ becomes slightly positive at 9 T. The reversed slope between 40 K and 30 K gives also a clue to the change of electronic structure.
Below 10 K, $\rho_{yx}$(B) increases reversely,
recovers the nearly linear behavior and holds the negative slope in the measured magnetic field range.
The changed slope and the non-linear field-dependence in $\rho_{yx}$(B) strongly suggest a multi-band system in W$_2$As$_3$, as reported
in other semimetals\cite{zhang2017electron,tafti2015LaSb,LuoY-WTe2Hall}.

For simplicity, we adopt a two-band model, with
one electron- and one hole-bands. In this model, the Hall conductivity ($\sigma_{xy}=\frac{\rho_{yx}}{\rho_{xx}^2+\rho_{yx}^2}$) reads as\cite{ziman1960},
\begin{equation}
\sigma_{xy}(B)= eB[\frac{n_h{\mu_h}^2}{1+{(\mu_hB)}^2}-\frac{n_e{\mu_e}^2}{1+{(\mu_eB)}^2}].
\label{Eq2}
\end{equation}
Particular attention was paid to this fitting (see \textbf{Experimental method}), and the derived results are summarized in Fig. 4(c).
At 100 K, electron-type carrier density $n_{e}$ dominates, while the mobilities of electron and hole are essentially the same, indicating that at this temperature the system can more or less be regarded as a single-band system, which is in contrast to the majority of hole-type carrier density in the previous report\cite{PhysRevB.98.115145ZAXu}. Lowering temperature, the $n_{e}$ only slightly decreases with $T$ (be noted the $\log$ scale of $n_{e,h}$) while the $n_{h}$ significantly increases in the temperature range 30-40 K. For $T \leq$ 25 K, $n_{e,h}$ tends to saturate and becomes comparable, confirming a nearly compensated electron-hole carrier densities in W$_2$As$_3$.
On the other hand, the mobility shows a large divergency below 40 K, with the value $\mu_e$ much larger than $\mu_h$ at low temperature,
consistent with the recovered negative slope in $\rho_{xy}$(B). All these features suggest a change of electronic structure at around 30-40 K.
The region 30-40 K from now on will be termed as $T_m$. Considering the rapid increase of $n_h$ near $T_m$, it is reasonable to assume that a new hole pocket shows up, and thus drastically modifies the electronic structure. Similar behaviors were observed in WTe$_2$ and MoTe$_2$ semimetals\cite{WTe2Lifshize,zhou2016hall}. This also gives an explanation to the change of ADMR near $T_m$, as the new hole pocket inevitably will change the anisotropy of Fermi surface.

As the extracted carrier mobilities ($\mu_{e,h}$) are in the range of 10$^{3-4}$ cm$^2$/Vs, we thus can calculate the value of $\omega_c \tau = \mu B \sim 3$ as $B$ =3 T, which is still far away from the high field regime($\omega_c\tau \gg 1$). In addition, the MR in a conventional metal should be saturated in the high field regime\cite{PRL94.166601}. However, in our sample such saturation behavior is not observed and the extremely large MR still follows the parabolic field-dependence even for the field up to 53 T. This implies that simply entering high field regime is not sufficient in this context, and the carrier compensation effect should be more important.

Additional evidences for such a temperature-induced Lifshitz transition can be provided by thermoelectric measurements. The thermopower is negative from 2-300K, confirming again electrons as the majority carrier. The zero-field thermopower divided by temperature ($S/T$) in Fig. 4(d) shows a non-monotonic temperature-dependence with a local maximum (in magnitude) at 40 K. Although this might be reminiscent of the phonon drag effect, the region that it starts to drop agrees surprisingly well with $T_m$, it, therefore, is more likely that such behavior is of electronic origin. Applying a magnetic field up to 9 T obviously enhances the magnitude of $S$ and suppresses the local maximum of thermopower towards the low temperatures (Fig. 3(d)). Such behavior is in contrast to the robust thermopower under magnetic fields caused by the phonon drag effect, as reported in FeGa$_3$\cite{PhysRevBFeGe3}. We also measured the transverse magneto-thermoelectric transport, viz. Nernst effect. Here we adopt a classical sign convention by which the Nernst effect caused by vortex motion is positive. It is apparently seen that above 50 K, the Nernst coefficient divided by temperature ($\upsilon/T$) is negligibly small. It is well known that for a single-band, nonsuperconducting and nonmagnetic metal, the Nernst signal is vanishing, due to the so-called Sondheimer cancellation\cite{Sondheimer}. Such a cancellation is violated below 40 K, and the magnitude of $\upsilon/T$ increases rapidly. A relevant example of large Nernst effect is the compensated semimetal Bi\cite{PRL.98.076603kamran} because of the equal density of electrons and holes in bands. The change in $\upsilon/T$ observed near $T_m$ in W$_2$As$_3$, again, is in agreement with a temperature-induced Lifshitz transition.

Turning now to the SdH quantum oscillations. Since systematic SdH studies have been reported recently in Ref.~\cite{PhysRevB.98.115145ZAXu}, we do not plan to get into too many details. After subtracting a smooth polynomial background from the total resistivity, the remaining part, $\rho$$-$$\langle \rho \rangle$, is shown in Fig.~5(a) as a function of $1/B$. The clear SdH oscillations were observed and the most of oscillation patterns consist of a main frequency $F$=337 T, in spite of some sub-features observed only for field above 50 T that is potentially due to the Zeeman effect. More SdH frequencies can be visible if we take Fast Fourier Transform (Data not shown here). Since $\rho_{xx}(B)$$\gg$$\rho_{yx}$ in W$_2$As$_3$, the maxima of the SdH oscillations in $\rho_{xx}(B)$ is assigned as integer Landau level (LL) indices $n$ (See e.g.\cite{Xiang_BiTeClSdH}), as shown in Fig.5(b). A linear extrapolation of $n$ versus 1/$B$ to the infinite field limit yields the intercept of -0.10(3). According to the Lifshitz-Kosevich formula, this intercept corresponds to $1/2-\Phi_B/2\pi-\delta$\cite{Murakawa-BiTeISdH}, where $\Phi_B$ is the Berry phase, and $-1/8\leq\delta\leq1/8$ is an additional phase shift stemming from the curvature of the Fermi surface topology. This means that a trivial (non-trivial) Berry phase gives rise to an intercept close to 0 (0.5). The intercept derived here, $\sim$ -0.10(3), falls between $\pm$1/8, and thus should point to a non-trivial topological feature. In short, from SdH-oscillation measurements we confirm W$_2$As$_3$ is a topological semimetal. This same conclusion was also drawn by Li et al\cite{PhysRevB.98.115145ZAXu}; there more Fermi sheets with different Berry phases were resolved.

The Fermi surface structure has also been reported in Li's work\cite{PhysRevB.98.115145ZAXu}, and several \textit{close} pockets, both electron- and hole-types, are shown by DFT calculations. For simplicity, a \textit{schematic} band structure of W$_2$As$_3$ can be proposed based on all these results, see Fig. 5(c). The compound should contain at least one topological band that has Dirac-like, linear dispersion. Considering a much higher electron mobility than hole mobility, it is more likely that this non-trivial band is of electron-type, as seen in Fig. 5(c). At high temperature, the chemical potential $\mu_F$ passes through this Dirac band only. The top of valence band is supposed to sit right below $\mu_F$. As temperature decreases, the chemical potential moves downward and when $T$=$T_m$$\sim$30-40 K, $\mu_F$ hits the top of a valence band, generating a great amount of holes that lead to the electron-hole compensation and large MR at low temperature. The emergence of the new hole pocket changes the anisotropy of Fermi surface, bearing the consequence for the modification of ADMR patterns. The thermoelectric properties also change correspondingly. To be more straightforward, a semi-quantitative density of states [DOS, $N(\varepsilon)$] can be drawn in this picture [cf. Fig.~5(d)]. Note that a three dimensional Dirac dispersion renders $N(\varepsilon)\propto\varepsilon^2$ (see the red curve in Fig.~5(d)). For $T$$\rightarrow$0, $\mu_F$ resides inside the valence band. A rule of thumb about the movement of chemical potential with raising temperature is that it shifts towards the side that has \textit{smaller} DOS, which in our case is to move up. This is because the electron density is
\begin{equation}
n_e=\int_{\mu_F}^{+\infty}N(\varepsilon)f(\varepsilon)d\varepsilon,
\label{Eq3}
\end{equation}
and the hole density is
\begin{equation}
n_h=\int_{-\infty}^{\mu_F}N(\varepsilon)[1-f(\varepsilon)]d\varepsilon,
\label{Eq4}
\end{equation}
where $f(\varepsilon)$=$\frac{1}{1+\exp{(\varepsilon-\mu_F)/T}}$ is Fermi-Dirac distribution function. The electron (for $\varepsilon$ above $\mu_F$) and hole (for $\varepsilon$ below $\mu_F$) distributions have been depicted by the pink and magenta areas in the right panel of Fig.~5(d), respectively. As $T$ increases, $f(\varepsilon)$ becomes more smeared-out, and therefore, $n_h$ increases faster than $n_e$; $\mu_F$ has to move up to compensate. According to the Mott relation\cite{ziman1960}, the thermopower is $S = \frac{\pi^2 k_B^2T}{3e\sigma_{xx}}\frac{\partial{\sigma_{xx}(\varepsilon)}}{\partial{\varepsilon}}|_{\varepsilon=\mu_F}$. Previous theoretic work by Blanter \textit{et al} predicted a maximum in thermopower when $\mu_F-\varepsilon_c$$\rightarrow$0 \cite{Blanter-ETT}, where $\varepsilon_c$ is the critical energy as a Lifshitz transition takes place. In our case this corresponds to the top of valence band. Such a maximum of thermopower (in magnitude) is indeed seen in our experiment as shown in Fig.~4(d). The realistic band structure of W$_2$As$_3$, although, is more complicated\cite{PhysRevB.98.115145ZAXu}, but such a temperature-induced Lifshitz transition seems rather likely.

Finally, it is worthwhile to mention that, one difficulty that we have to confront when comparing the experimental results to the theoretical calculations is that we do not know exactly where the chemical potential pins to. DFT calculation loses its power in this context because it varies in realistic samples, depending on the content of nonstoichiometry and disorder. Angle resolved photoemission spectroscopy (ARPES) is required to further clarify the problem, as it is the case in WTe$_2$\cite{WTe2Lifshize}.

In conclusion, compared with the previous study in W$_2$As$_3$\cite{PhysRevB.98.115145ZAXu}, besides the large unsaturated MR, more importantly, we first find that the ADMR of W$_2$As$_3$ depends strongly on temperature, and undergoes an obvious deformation from the horizontal dumbbell-like shape above 40 K to the vertical lotus-like pattern below 30 K. Such feature implies a dramatic reconstruction of Fermi surface.
Hall effect measurements indicate an abnormal increase in hole carrier density below 40 K, around which thermopower exhibits a large reversal while Nernst coefficient increases abnormally. All of these experimental results point to an interesting Lifshitz transition occurring between 30-40 K, which in turn leads to the compensated electron-hole carrier density, as well as the large MR in W$_2$As$_3$. Our works not only pave an important path to investigate the origin of large MR, but also suggest that the simple ADMR measurement will be applicable in detecting a Fermi-surface change of a variety of quantum states at a low expense but high convenience, especially when a direct Fermi-surface measurement is not possible.

\section{\label{sec:level1}Experimental method}

Single crystals W$_2$As$_3$ were grown through chemical vapor transport reaction
using iodine as transport agent. Polycrystalline samples of W$_2$As$_3$ were first synthesized by
solid state reaction using high purified Tungsten powders and
Arsenic powders in a sealed quartz tube at 973 K for three days.
The final powders together with a transport agent iodine concentration of 10 mg$/cm^3$ were mixed thoroughly and then sealed
in a quartz tube with a low vacuum-pressure of $\leq$ 10$^{-3}$ Pa. The tube was heated in
a horizontal tube furnace with a temperature gradient of 120 $^{\circ}{\rm C}$
between 1120 $^{\circ}{\rm C}$ -1000 $^{\circ}{\rm C}$ for 1-2 weeks.
The obtained high-quality single crystals are shiny and black with the hexagonal-shape.

X-ray diffraction patterns were performed using a D/Max-rA diffractometer with CuK$_{\alpha}$
radiation and a graphite monochromator at the room temperature. The
single crystal X-ray diffraction determines the crystal grown
orientation. The energy dispersive X-ray (EDX) spectroscopy was employed to verify the composition of the crystals,
yielding a fairly homogenous stoichiometric ratio. No iodine impurity can be detected in these single crystals.
Electrical transport measurements were performed using the standard
six-terminal method with electric current flowing in parallel to the $\mathbf{b}$-axis. Ohmic contacts were carefully prepared
on the crystal in a Hall-bar geometry. The low contact resistance was obtained after annealing the silver paste at 573 K for an hour.
Thermoelectric properties were measured by a steady-state technique and a pair of differential type-E thermocouples
was used to measure the temperature gradient. The field dependence of thermocouple was pre-calibrated carefully. All these measurements were carried out in a commercial Quantum Design PPMS-9 system
with a rotator insert. The resistivity measurements under a pulsed magnetic field up to 53 T
were performed in Wuhan National High Magnetic Field Center.

In Figure 4, we performed the fitting of two band model in Hall data by two ways. First, we started to independently fit the Hall data at 100 K, and
yielded the carrier densities and mobilities. For the next temperature 70 K, we employed the values at 100 K as the initial parameters, and thus gets the reliable fitting for 80 K. The same procedure keeps on until we finished the last temperature 2 K. Second, in order to get a further check, we also \textit{independently} started the fitting at 2 K, and gradually moved back to 100 K. The obtained two sets of parameters by this opposite way are in good agreement. Such self-consistent fitting method makes our fitting results reliable.

\section*{Data Availability}

All data generated or analysed during this study are included in the present manuscript.

\section*{Acknowledgements}
This research was supported in part by the NSF of China (under
Grants No. 11874136, U1932155, U1732274). Yu-Ke Li was supported by an open
program from Wuhan National High Magnetic Field Center (2016KF03).
Jie Cheng was supported by the General Program of Natural Science Foundation of Jiangsu Province of China (No. BK20171440). Y. Luo acknowledges the 1000 Youth Talents Plan of China.

\section*{Author contributions}

J. Wang and H. Yang contributed equally to this work. Y. Li designed the research.
J. Wang synthesized the samples. J. Wang and H. Yang performed the electronic and thermal transport measurements.
L. Ding performed the resistivity measurements under high magnetic fields in Wuhan.
W. You performed the XRD measurements and analyzed the structure parameters.
C. Xi and J. Cheng assisted the measurements.
Z. Shi, Y. Luo, Z. Zhu, C. Cao, J. Dai, Y. Li, and M. Tian discussed the data, interpreted the results. Y. Luo, J. Dai and Y. Li wrote the paper.

\section*{Additional information}
Competing financial interests: The authors declare no competing Financial Interests or Non-Financial Interests.
Correspondence and requests for materials should be addressed to
Y. Li (yklee@hznu.edu.cn), J. Dai (daijh@hznu.edu.cn), Y. Luo (mpzslyk@gmail.com), or M. Tian (tianml@hmfl.ac.cn).

\clearpage

\begin{figure}[htp]
\includegraphics[width=12cm]{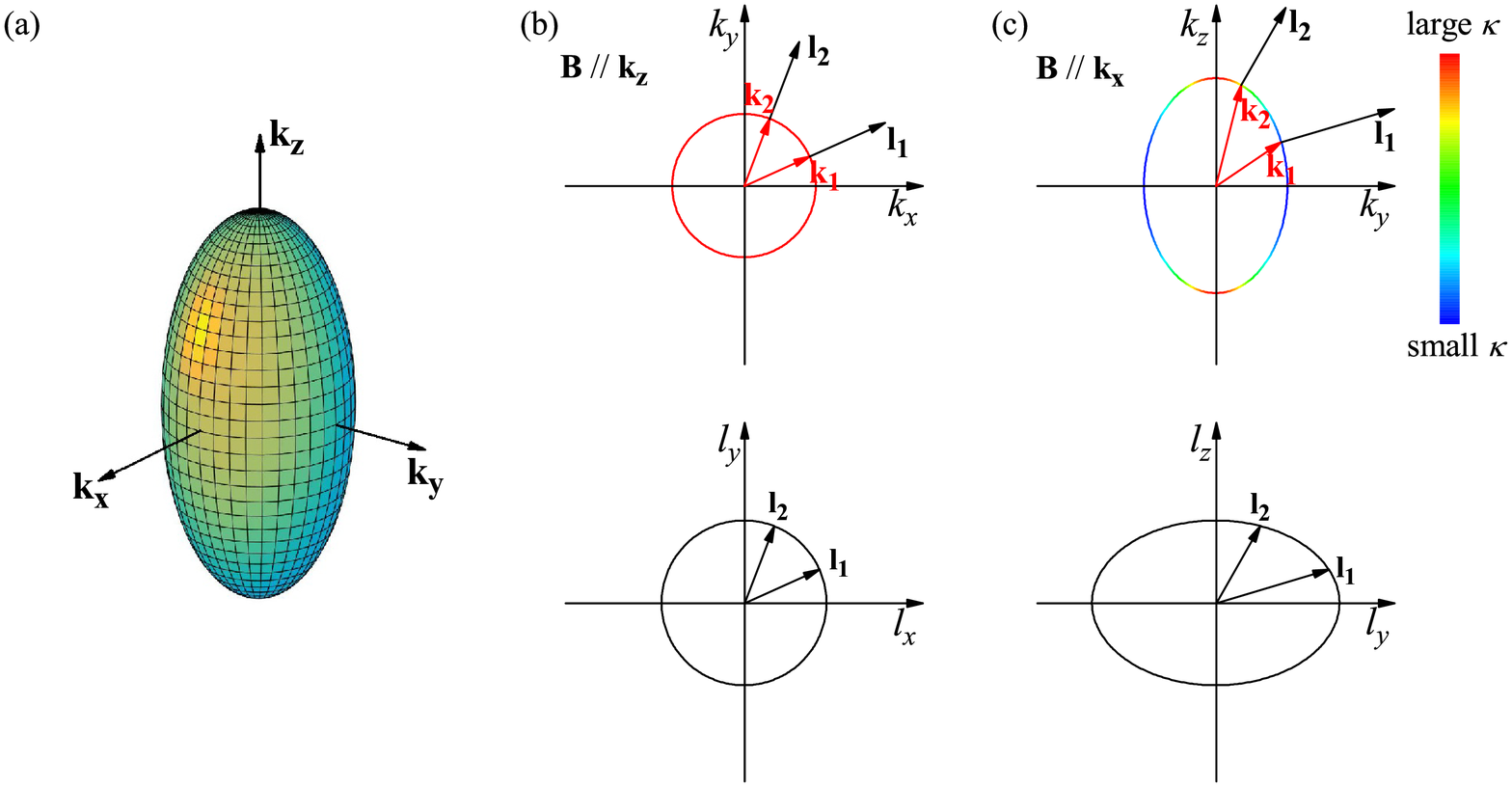}
\caption{\label{Fig1}
Magnetoresistance and Fermi surface geometry. (a) An ellipsoid Fermi surface whose semiaxis is elongated along $\mathbf{k_z}$. (b) For $\mathbf{B}\parallel\mathbf{k_z}$, the Fermi surface cross-section is a circle (top panel), and so is the scattering path length $\mathbf{l}$ curve (bottom panel). (c) For $\mathbf{B}\parallel\mathbf{k_x}$, the Fermi surface cross-section and mapped out $\mathbf{l}$ curve; the Fermi surface is colored with the local curvature $\kappa$. Note that $\mathbf{l}(\mathbf{k})=\mathbf{v_k}\tau$, and $\mathbf{v_k}=\frac{1}{\hbar}\nabla_{\mathbf{k}}\varepsilon_{\mathbf{k}}$. }
\end{figure}

\begin{figure}
\includegraphics[width=12cm]{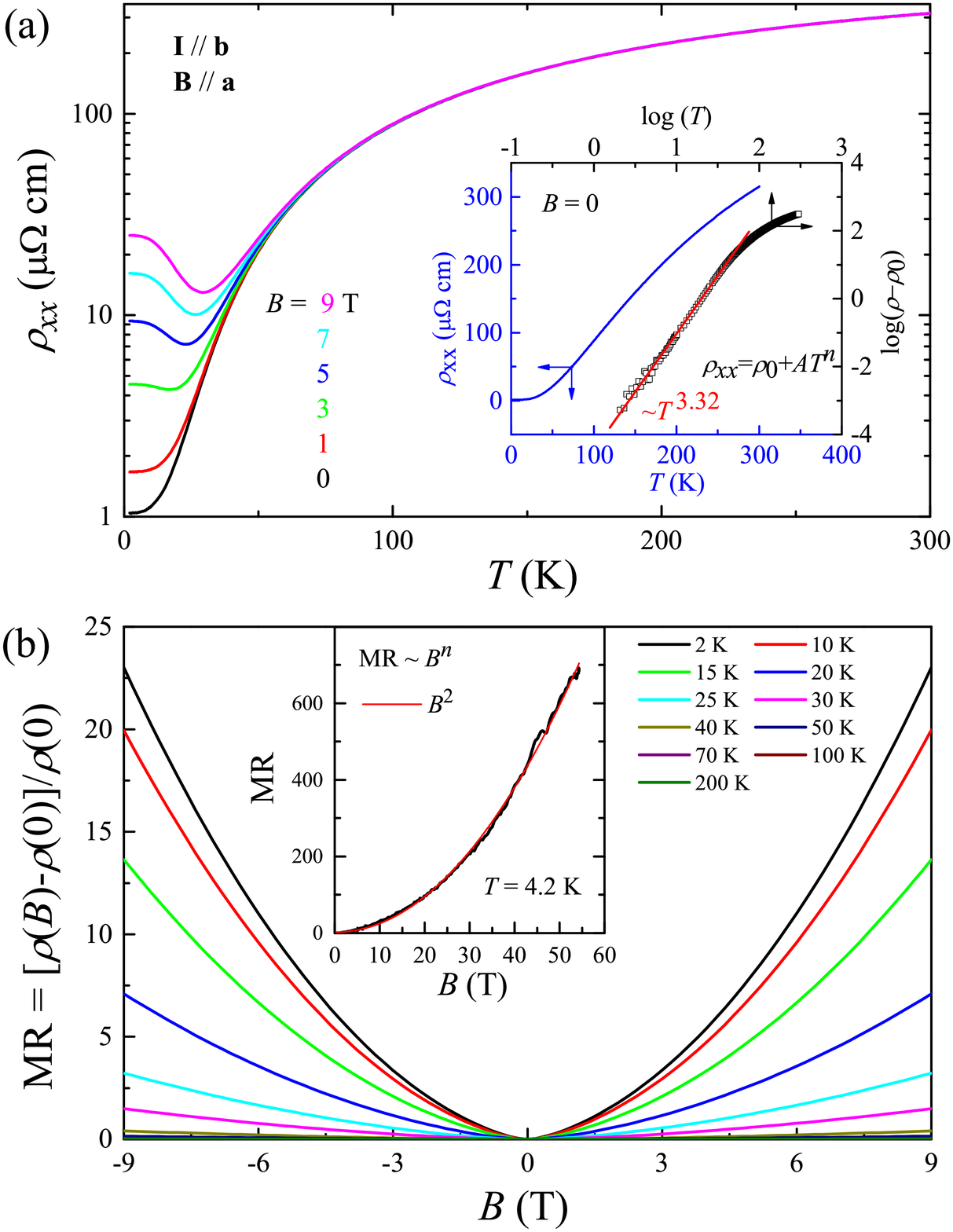}
\caption{\label{Fig2}
Magnetoresistance of W$_2$As$_3$. (a) Temperature dependence of resistivity of W$_2$As$_3$ under several magnetic fields ($B$=0, 1, 3, 5, 7, 9 T.)
perpendicular to the current. Inset shows $\rho_{xx}(T)$ for $B$=0 (bottom-left frame) and $\log(\rho-\rho_0)$ vs. $\log(T)$ (top-right frame), the latter of which gives rise to $\rho(T)\propto T^{3.32}$ at low temperatures.
(b) MR as a function of magnetic fields at different temperatures below 200 K. The inset shows MR vs. $B$/$\rho(0)$ at 4.2 K for field up to 53 T. }
\end{figure}

\begin{figure*}
\includegraphics[angle=0,width=18.cm,clip]{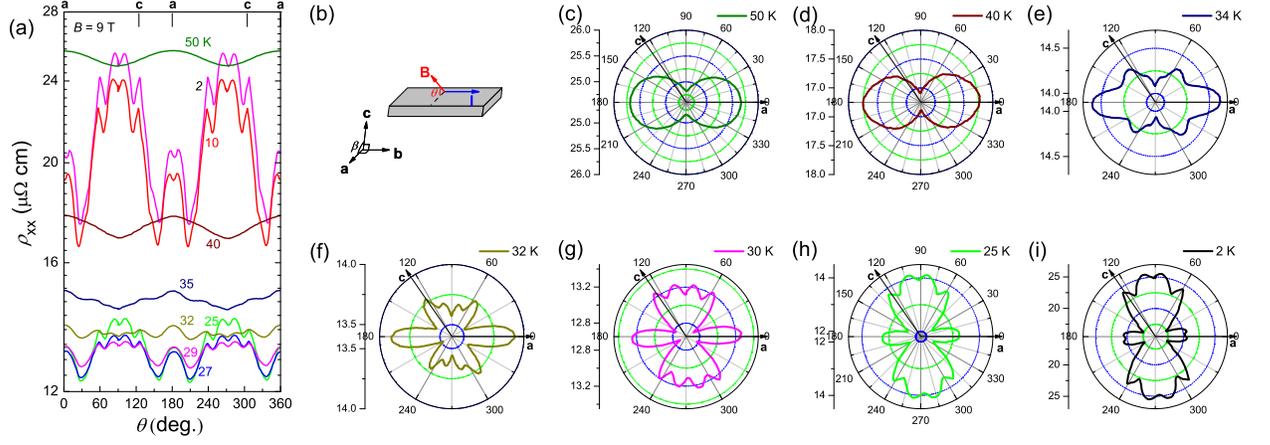}
\label{Fig3}
\caption{
Angular dependent magnetoresistance of W$_2$As$_3$. (a) Field-rotation resistivity at different temperatures. The applied magnetic field is fixed at 9 T, and the rotation is about $\mathbf{b}$-axis which is parallel with electric current $\mathbf{I}$, cf. panel (b). $\beta=124.7\textordmasculine$ characterizes the angle between $\mathbf{a}$ and $\mathbf{c}$. (c)-(g) Several representative patterns of ADMR displayed in polar plots.}
\end{figure*}

\begin{figure*}
\includegraphics[angle=0,width=16cm,clip]{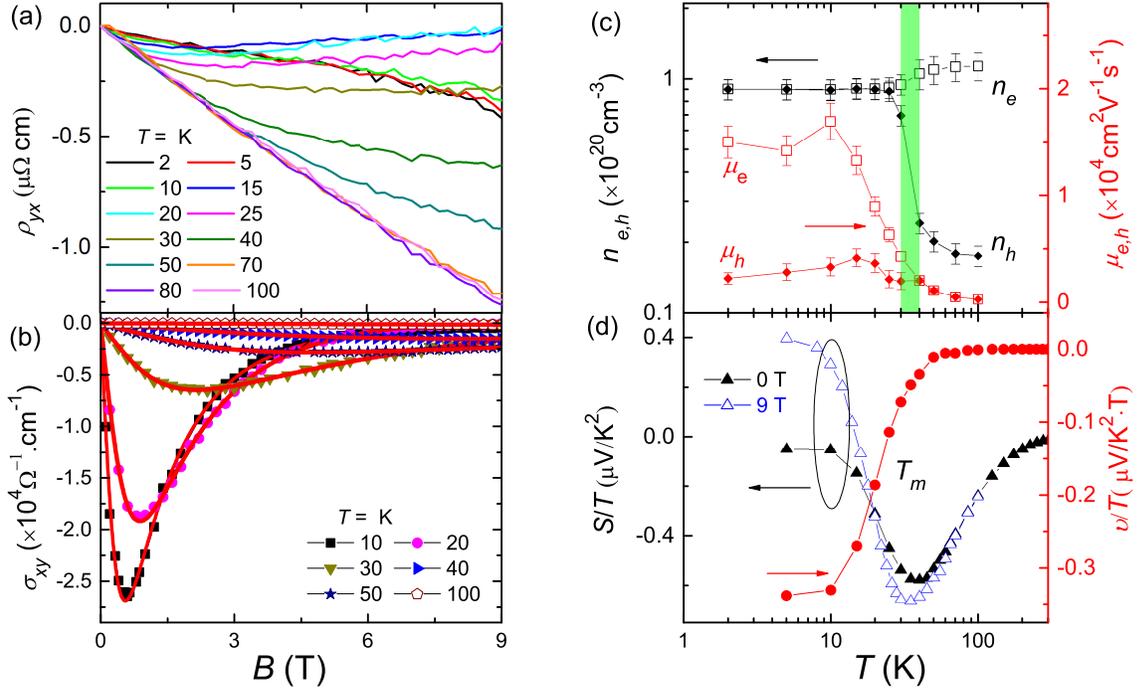}
\label{Fig4}
\caption{
Hall and thermoelectric effects of W$_2$As$_3$.
(a) Magnetic field dependence of Hall resistivity at several representative temperatures.
(b) Hall conductivity vs. magnetic fields at selected temperatures. The red solid lines are the fitting curves using two-band model.
(c) Carrier densities (left frame) and mobilities (right frame) as functions of temperature. The open (solid) symbols denote electron (hole).
(d) The thermopower $S/T$ at zero field and 9 T and the Nernst coefficient $\upsilon/T$ vs. $T$. The region 30$<$$T$$<40$ K shaded with green colour signify the temperature $T_m$ around which the Lifshitz transition takes place.}
\end{figure*}

\begin{figure*}
\includegraphics[angle=0,width=16cm,clip]{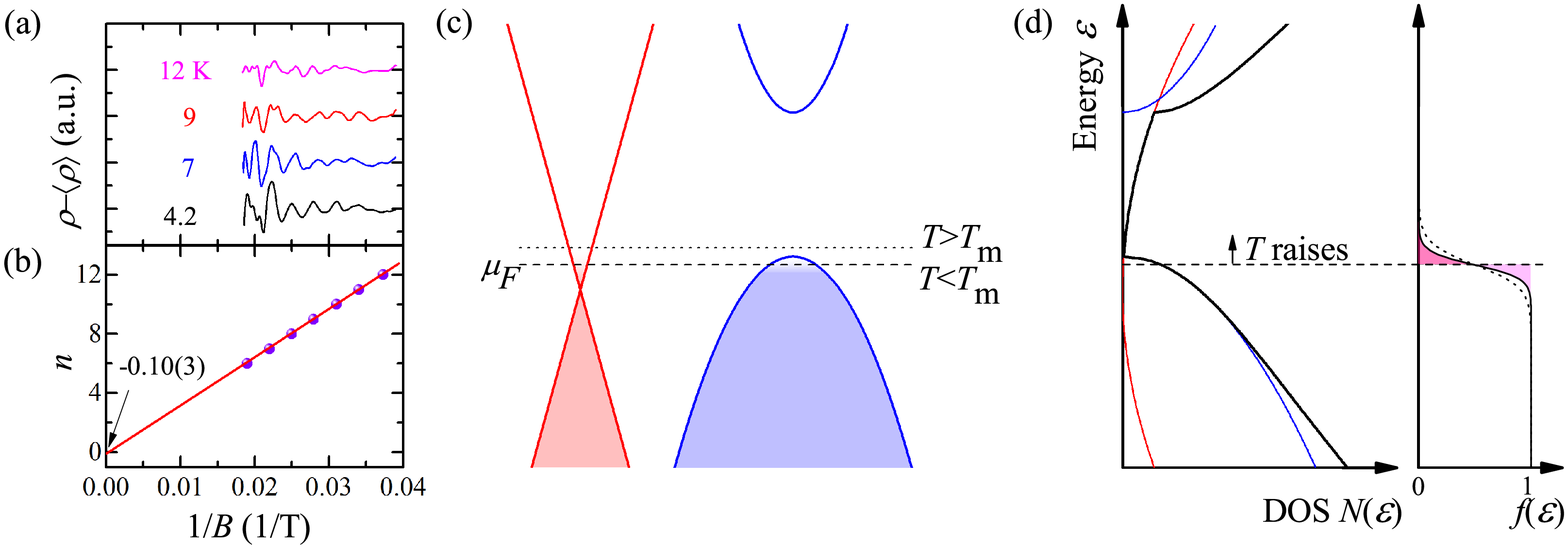}
\label{Fig5}
\caption{
Quantum oscillations and band structure.
(a) SdH oscillations observed in resistivity, displayed as a function of $1/B$.
(b) Landau fan diagram. The linear extrapolation to the infinite field limit results in the intercept of $\sim -$0.10(3).
(c) The low-energy band structure consists of both topologically non-trivial (red) and trivial (blue) bands. Above $T_m$, the chemical potential ($\mu_F$) is supposed to pass through the Dirac-like non-trivial band; $\mu_F$ shifts downward upon cooling and cuts the trivial valence band at $T_m$. (d) Schematic diagram of density of states (DOS). The colors are ibid, and the black line stands for the total DOS. The right panel shows Fermi-Dirac distribution function. The electron and hole distributions are depicted by the pink and magenta areas. The movement of $\mu_F$ as $T$ raises is depicted by the arrow.}
\end{figure*}

\end{document}